\documentclass[iop]{emulateapj}
\usepackage{epstopdf}
\usepackage{subfigure}
\usepackage{amsmath}

\usepackage{longtable}

\shorttitle{Alfv\'enic Wave Heating}
\shortauthors{Reep \& Russell}

\begin{document}

\title{Alfv\'enic Wave Heating of the Upper Chromosphere in Flares}

\author{J. W. Reep}
\affil{National Research Council Post-Doc Program, Naval Research Laboratory, Washington, DC 20375 USA}
\email{jeffrey.reep.ctr@nrl.navy.mil}
\and
\author{A.J.B. Russell}
\affil{Division of Mathematics, University of Dundee, Nethergate, Dundee, DD1 4HN, Scotland, UK}
\email{arussell@maths.dundee.ac.uk}

\begin{abstract}
We have developed a numerical model of flare heating due to the dissipation of Alfv\'enic waves propagating from the corona to the chromosphere.  With this model, we present an investigation of the key parameters of these waves on the energy transport, heating, and subsequent dynamics.  For sufficiently high frequencies and perpendicular wave numbers, the waves dissipate significantly in the upper chromosphere, strongly heating it to flare temperatures.  This heating can then drive strong chromospheric evaporation, bringing hot and dense plasma to the corona.  We therefore find three important conclusions: (1) Alfv\'enic waves, propagating from the corona to the chromosphere, are capable of heating the upper chromosphere and the corona, (2) the atmospheric response to heating due to the dissipation of Alfv\'enic waves can be strikingly similar to heating by an electron beam, and (3) this heating can produce explosive evaporation.  
\end{abstract}

\keywords{Sun: chromosphere, Sun: corona, Sun: flares, Sun: atmosphere, Sun: transition region, waves}

\section{Introduction}
\label{sec:introduction}

The standard model of solar flares, referred to as the CSHKP model \citep{carmichael1964,sturrock1966,hirayama1974,kopp1976}, explains many observational features of flares, assuming that they are driven by magnetic reconnection.  After the reconnection event triggers, between approximately $10^{30}$--$10^{33}$\,erg is released into the plasma, driving intense heating and brightening across the electromagnetic spectrum \citep{fletcher2011}.  It is not clear, however, how that energy is partitioned between {\it in situ} heating of the corona, particle acceleration, and wave generation, nor to what extent the observable features of a flare depend on the balance between different types of coronal energy transport.  

The collisional thick-target model (CTTM) \citep{brown1971} assumes that the released energy goes into acceleration of coronal particles, primarily electrons, to extremely high energies.  Down-going particles stream through the corona, eventually colliding with the much denser chromosphere where they lose energy through Coulomb collisions.  This energy loss in turn heats the chromosphere, driving evaporation into the corona, heating the loop and producing the sharp rises in intensities in the soft X-rays and extreme ultraviolet.  Observations of hard X-ray (HXR) bursts in flares show without question the braking of high energy electrons in the chromosphere, with as many as $10^{36}$\,s$^{-1}$ inferred for large flares ({\it e.g.} \citealt{holman2003}), however, these observations do not exclude the possibility of additional energy transport by means other than runaway particles.  Indeed, the flux of high energy electrons braking in the chromosphere presents several challenges to the classic CTTM, discussed by \citet{fletcher2008}, \citet{brown2009}, \citet{krucker2011}, and \citet{melrose2014}.  It has been suggested that these issues could potentially be resolved if some of the flare energy were transported through the corona by waves, and used to either accelerate electrons in higher density regions \citep{fletcher2008} or reaccelerate energetic particles \citep{brown2009, varady2014}.

In this paper, we examine Alfv\'enic waves as a heating mechanism that may act separately or in addition to the thick-target model.  Magnetohydrodynamic waves \citep{alfven1942} are observed ubiquitously in the corona \citep{tomczyk2007,mcintosh2011}, and often considered a leading candidate to explain coronal heating \citep{klimchuk2006} and the FIP effect \citep{laming2004, laming2015}.  For flares, Alfv\'en and guided fast waves produced during reconnection can deliver concentrated Poynting flux to the chromosphere \citep{birn2009, russell2013b}, where they damp in the cool, partially ionized plasma \citep{depontieu2001, khodachenko2004,soler2015}.  Simulations of magnetic reconnection show that Alfv\'en waves carry a large fraction of the released energy ($> 30\%$) in low $\beta$ plasma \citep{kigure2010}.  Previous studies by \citet{emslie1982} and \citet{russell2013} have shown that energy transport by Alfv\'enic waves can explain temperature minimum heating observed in solar flares, where temperature rises approximately 100\,K \citep{machado1978,emslie1979}.  Here, we emphasize the ability of Alfv\'enic waves to heat the upper chromosphere.
 
Following the formalism of \citet{russell2013} and references therein, we combine a hydrodynamic model with energy transport through Alfv\'enic waves, whereby the waves propagate from the reconnection site in the corona towards the chromosphere.  We present results from simulations that vary the wave parameters in order to show directly that not only can waves heat the temperature minimum region, but they can also heat the upper chromosphere.  Further, the heating can appear extremely similar to an electron beam, and can drive explosive evaporation.  We comment on the implications for the interpretation of observations of solar flares.

\section{Heating by Alfv\'enic Waves}
\label{sec:theory}

\citet{emslie1982} developed a model of Alfv\'en wave heating to explain the observed temperature minimum region heating observed in solar flares \citep{machado1978}.  Using a WKB approximation, they derive an expression for the period-averaged Poynting flux as a function of distance along a magnetic flux tube with a finite resistivity, where decreases in the Poynting flux are assumed to heat the plasma.  We adopt their model, although we derive the WKB result using an ambipolar resistivity instead of the classical resistivity to better account for ion-neutral collisions in the chromosphere, which are vitally important for wave damping \citep{piddington1956, leake2014}.  

The Alfv\'enic waves are injected at the top of the model with period-averaged Poynting flux $S_{0}$, which damps according to Equation 2.18 of \citet{emslie1982}:
\begin{equation}
S(z) = S_{0} \exp{\Bigg(- \int_{0}^{z} \frac{dz'}{L_{D}(z')}\Bigg)}
\end{equation}
\noindent where $z$ is the curvilinear coordinate along the loop and $L_{D}(z)$ is an effective damping length, given by
\begin{eqnarray}
L_{D}(z) &=& \Bigg(\frac{1}{L_{\perp}(z)} + \frac{1}{L_{\parallel}(z)} \Bigg)^{-1} \nonumber \\
	&=& \Bigg( \frac{\eta_{\parallel} k_{x}^{2} c^{2}}{4 \pi v_{A}} + \frac{\eta_{\perp} \omega^{2} c^{2}}{4 \pi v_{A}^{3}} \Bigg)^{-1} \nonumber \\
	&=& \frac{4 \pi v_{A}^{3}}{c^{2}(\eta_{\parallel}k_{x}^{2} v_{A}^{2} + \eta_{\perp} \omega^{2})}
\label{dampinglength}
\end{eqnarray}
\noindent where $v_{A}$ is the local Alfv\'en speed, $c$ the speed of light, $k_{x}$ the perpendicular wave number, $\omega$ the angular frequency, and $\eta_{\perp}$ and $\eta_{\parallel}$ the perpendicular and parallel resistivities.  The perpendicular wave damping term includes Cowling resistivity ({\it e.g.} \citealt{soler2013}), such that 
\begin{eqnarray}
\eta_{\perp} &=& \eta_{\parallel} + \eta_{C} \\
	&=& \frac{m_{e} (\nu_{ei} + \nu_{en})}{n_{e} e^{2}} + \frac{B^{2} \rho_{n}}{c^{2} \nu_{ni} \rho_{t}^{2} (1 + \xi^{2} \theta^{2} )} \nonumber
\end{eqnarray} 
\noindent We use the subscripts e, i, n, and t to refer to electrons, ions, neutrals, and total, respectively.  Each $\nu$ refers to a collision frequency (the equations are listed in \citealt{russell2013}, and note a typo in that work: $\nu_{ni}$ should scale as $T_{i}^{1/2}$), while $n$ is the number density, $\rho$ the mass density, $\xi$ the ionization fraction of hydrogen ($\xi = \rho_{i}/\rho_{t}$), and $\theta = \omega/\nu_{ni}$.  The local Alfv\'en speed is modified by the presence of neutrals:    
\begin{equation}
v_{A}(z) = \frac{B}{\sqrt{4\pi \rho_{t}}}\Big(\frac{1 + \xi \theta^{2}}{1 + \xi^{2} \theta^{2}}\Big)^{1/2}
\end{equation}
which reduces to the standard expression in the high frequency limit and in the fully ionized case.  The wave damping is due to collisions between the different species, which decreases the wave amplitude and heats the local plasma as the waves propagate along the flux tube.  As in \citet{emslie1982}, the heating term $Q(z)$ is calculated from the decreasing Poynting flux as 
\begin{equation}
Q(z) = - \frac{dS}{dz}
\label{heatingterm}
\end{equation}
\noindent where the friction due to the Cowling term heats the ions and the rest heats the electrons.

The WKB model is reasonably easy to implement within existing codes, accommodates a wide range of wave properties, has a small computational overhead, and is an accurate approximation when used appropriately.  The main restriction is that the derivation assumes that the parallel wavelength, $2\pi v_{A}(z)/\omega$, is less than or similar to the gradient length scale of $v_{A}(z)$, making reflection negligible.  This is a good approximation for Alfv\'en waves with frequencies of 1\,Hz or higher once they are in the chromosphere.  Since the model does not account for wave reflection at the transition region, which can be substantial \citep{emslie1982,russell2013}, we set $S_{0}$ to produce a suitable Poynting flux immediately below the transition region so that appropriate heating rates are obtained for the chromosphere and accept that the model underestimates the coronal heating associated with a given level of chromospheric heating.  

\section{Numerical Model}
\label{sec:modeling} 

We have implemented the Alfv\'en heating model outlined in Section \ref{sec:theory} in the state-of-the-art Hydrodynamics and Radiation Code (HYDRAD; \citealt{bradshaw2003}), which solves the one-dimensional equations describing conservation of mass, momentum, and energy for a two-fluid plasma confined to an isolated magnetic flux tube (the current version's equations are detailed in \citealt{bradshaw2013}).  The code does not evolve the magnetic field, so that rather than evaluating the propagation and damping of Alfv\'enic waves, the code emulates wave heating with the WKB approximation.  

The model chromosphere is based on the VAL C model \citep{vernazza1981}, along with the approximation to optically thick radiative losses prescribed by \citet{carlsson2012}, and the effects of neutrals as detailed in \citet{reep2013}.  The electron beam heating model used in Section \ref{sec:results} is based on \citet{emslie1978}.  For all of the simulations here, we employ a full loop with length $2L = 60$\,Mm, initially tenuous and in hydrostatic equilibrium, semi-circular and oriented vertically, and assume that the heating is symmetric about the apex.  

An important feature of HYDRAD is its ability to solve for non-equilibrium ionization states \citep{bradshaw2003}.  Since the resistivities depend on collisions between ions and electrons as well as between ions and neutrals, it is critically important to properly treat the ionization state of the plasma.  In particular, if the plasma is rapidly heated, the ionization state may lag behind the actual temperature.  HYDRAD uses ionization and recombination rates taken from the CHIANTI v.8 database \citep{dere1997,delzanna2015} to solve for the ionization fraction of a given element.  We treat radiative losses using a full calculation with CHIANTI.

Equation \ref{heatingterm} is taken as the heating term in the energy equation.  The ionization balance is solved with the following equation \citep{bradshaw2003}: 
\begin{equation}
\frac{\partial Y_{i}}{\partial t} + \frac{\partial}{\partial s}(Y_{i} v) = n_{e} (I_{i-1} Y_{i-1} + R_{i} Y_{i+1} - I_{i} Y_{i} - R_{i-1} Y{i})
\end{equation}
\noindent where $Y_{i}$ is the fractional population of an ionization state $i$ of element $Y$, and $I_{i}$ and $R_{i}$ are the ionization and recombination rates from and to $i$ (respectively).  Solving this equation for hydrogen gives the fractional population of ion and neutral densities in the chromosphere that determine the damping length. 

To study the effects of chromospheric heating by Alfv\'enic waves, we have run $24$ numerical experiments.  These simulations cover a wide range of possible values, and allow for systematic investigation of wave heating.  We vary the wave frequency $f = \omega / 2 \pi $ between [1, 10, 100]\,Hz and the perpendicular wave number at the loop apex $k_{x,a}$ between [0, $10^{-5}$, $10^{-4}$, $4 \times 10^{-4}$]\,cm$^{-1}$.  For evaluation of $L_{D}$ from Equation \ref{dampinglength}, we assume that the magnetic field has a photospheric value $B_{0} = 1000$ G, decreasing along the flux tube with the pressure as $B(z) = B_{0} (\frac{P(z)}{P_{0}})^{0.139}$ (as in \citealt{russell2013}), which is constant in time.  Since the density and magnetic field vary with position, the Alfv\'en speed $v_{A}$ also varies.  We also adopt two different dependencies for $k_x$ as a function of position: with $k_{x}(z) = k_{x,a} (\frac{B(z)}{B_{a}})$ (linear in $B$) for magnetic expansion in one dimension as in an arcade geometry, and $k_{x}(z) = k_{x,a} \sqrt{\frac{B(z)}{B_{a}}}$ (as the square root) for magnetic expansion in two dimensions as in a flux tube that expands radially with height.  We do not scale the flux density with the changing cross-sectional area implied from the expansion of $B(z)$.

\section{Results}
\label{sec:results}

\begin{figure*}
\begin{minipage}[b]{0.33\linewidth}
\centering
\includegraphics[width=\textwidth]{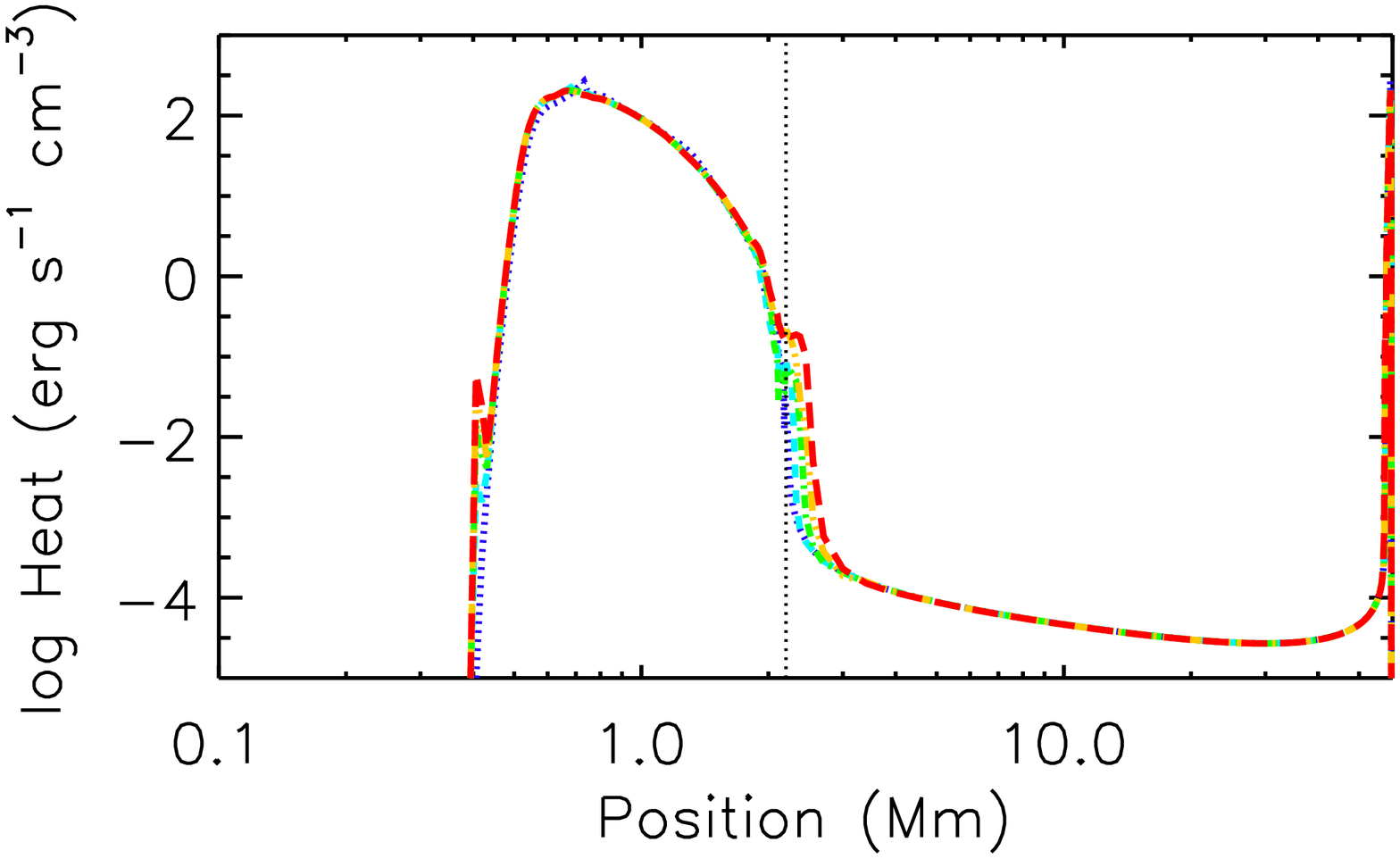}
\end{minipage}
\begin{minipage}[b]{0.33\linewidth}
\centering
\includegraphics[width=\textwidth]{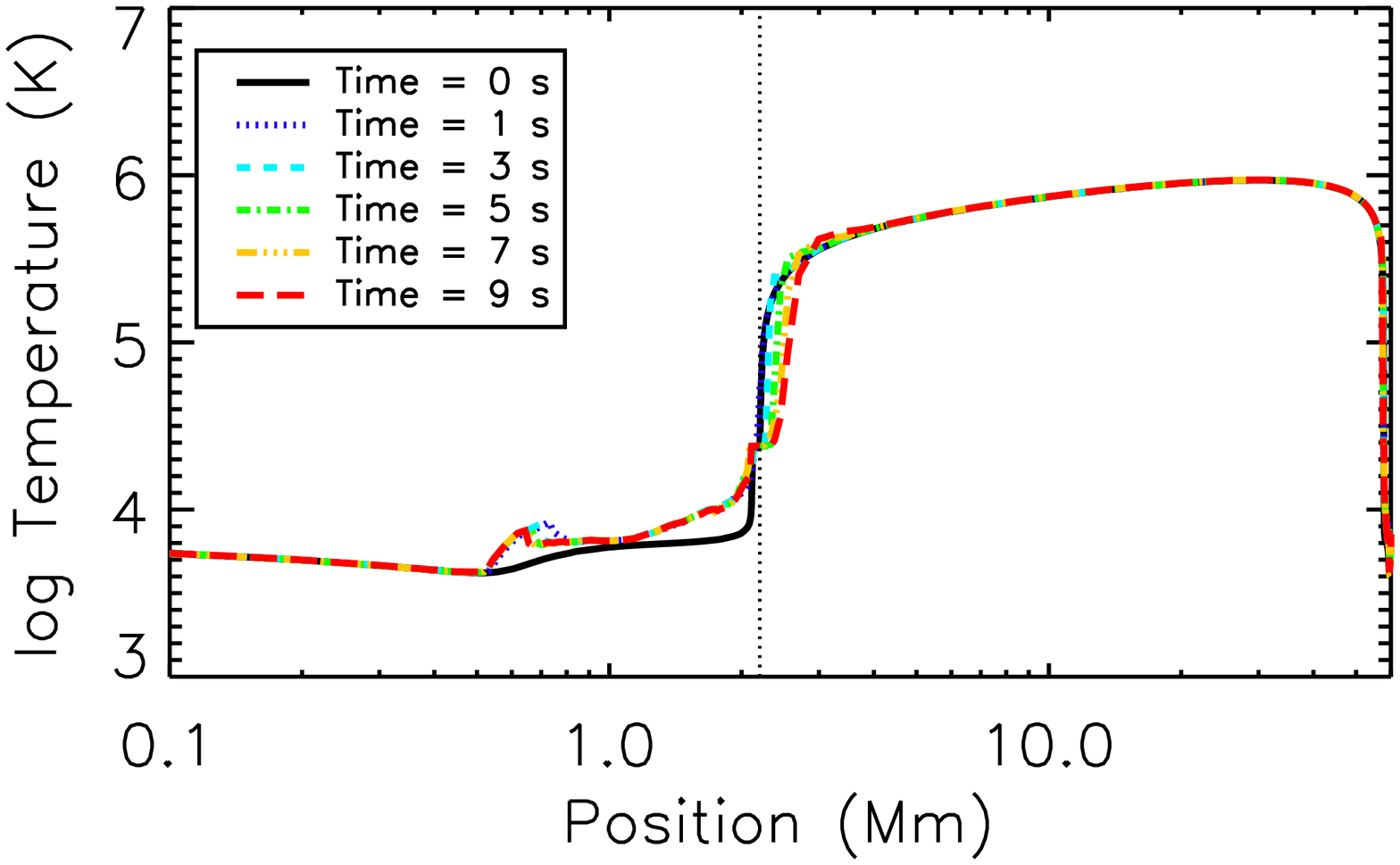}
\end{minipage}
\begin{minipage}[b]{0.33\linewidth}
\centering
\includegraphics[width=\textwidth]{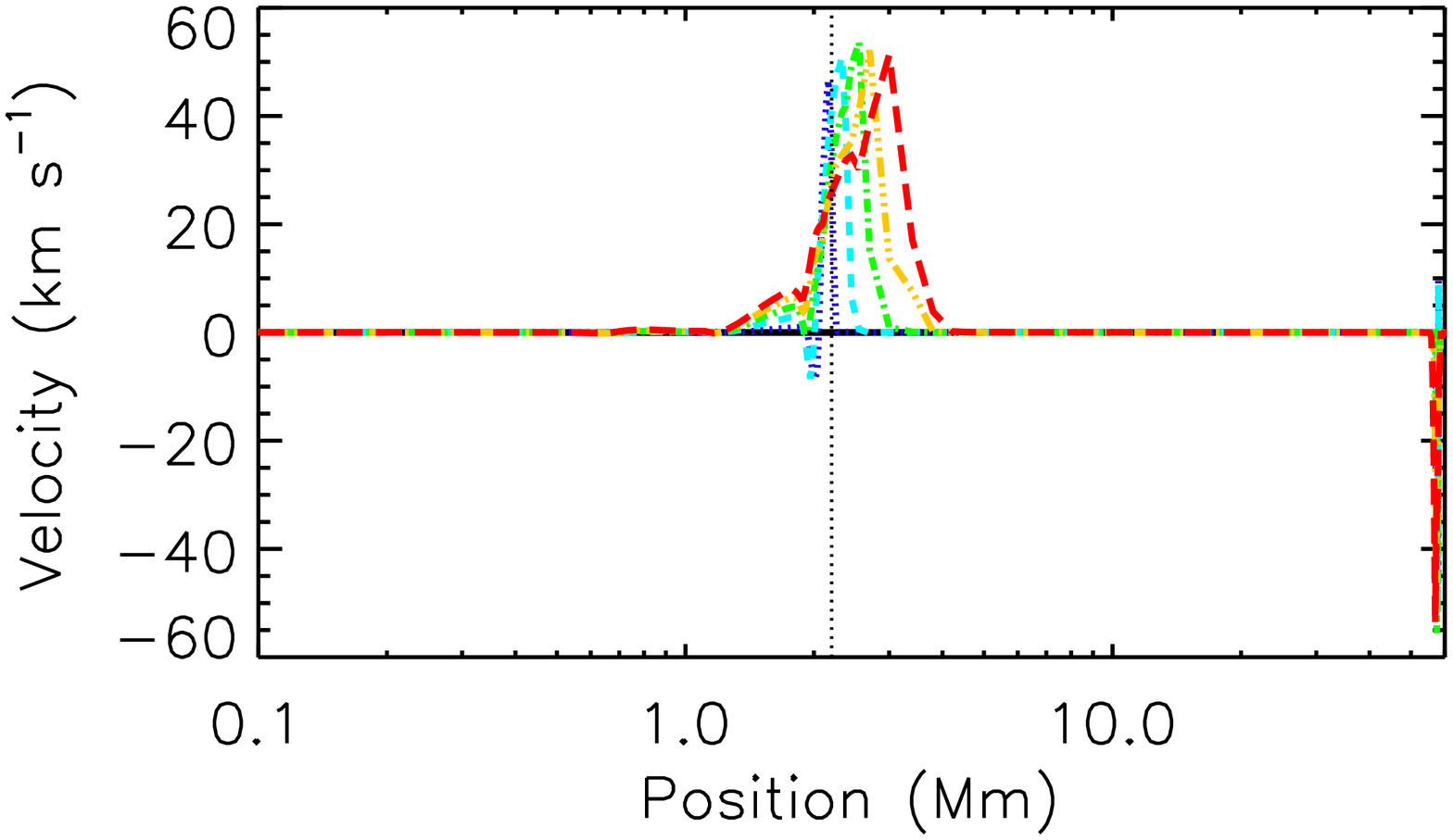}
\end{minipage}
\begin{minipage}[b]{0.33\linewidth}
\centering
\includegraphics[width=\textwidth]{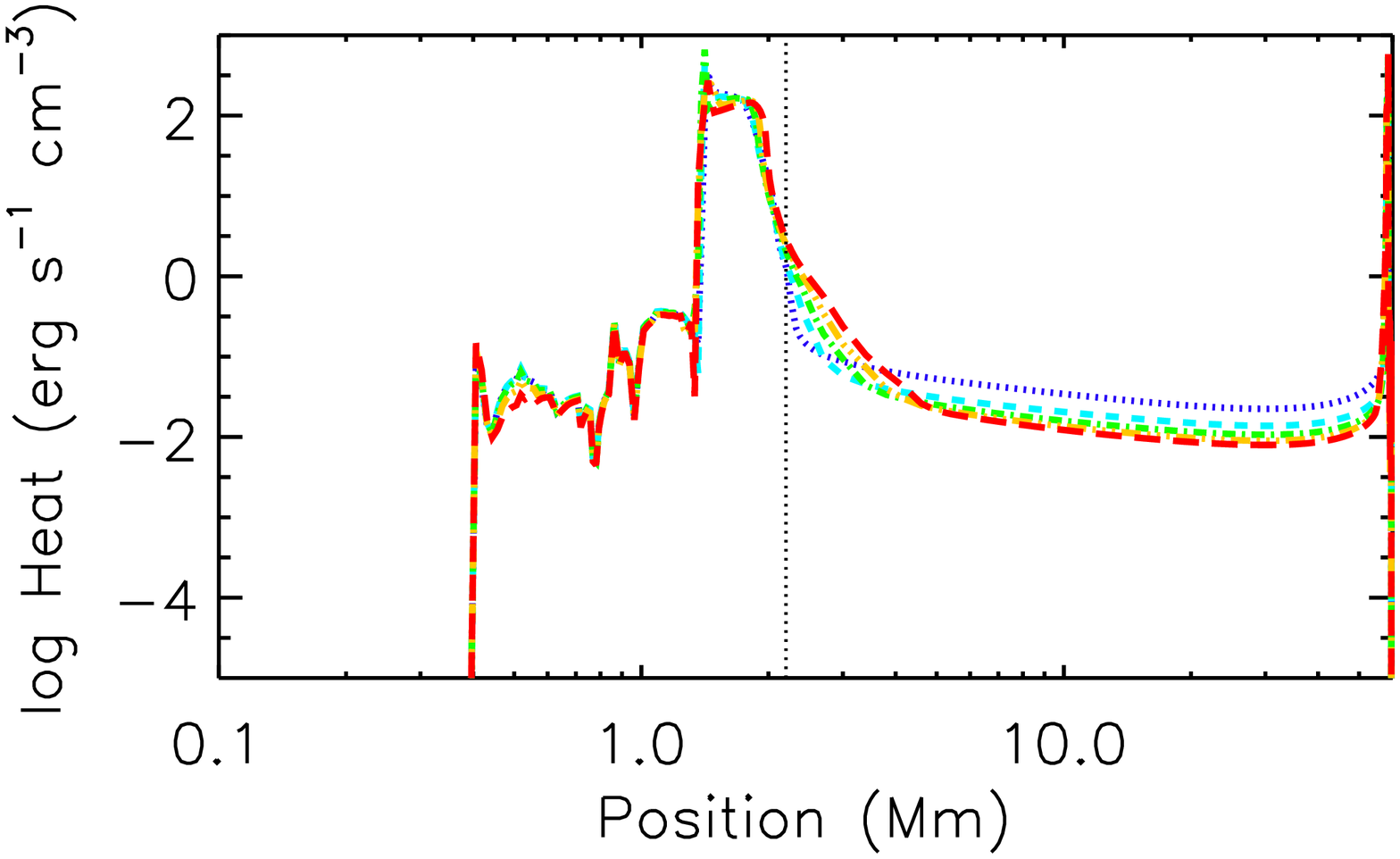}
\end{minipage}
\begin{minipage}[b]{0.33\linewidth}
\centering
\includegraphics[width=\textwidth]{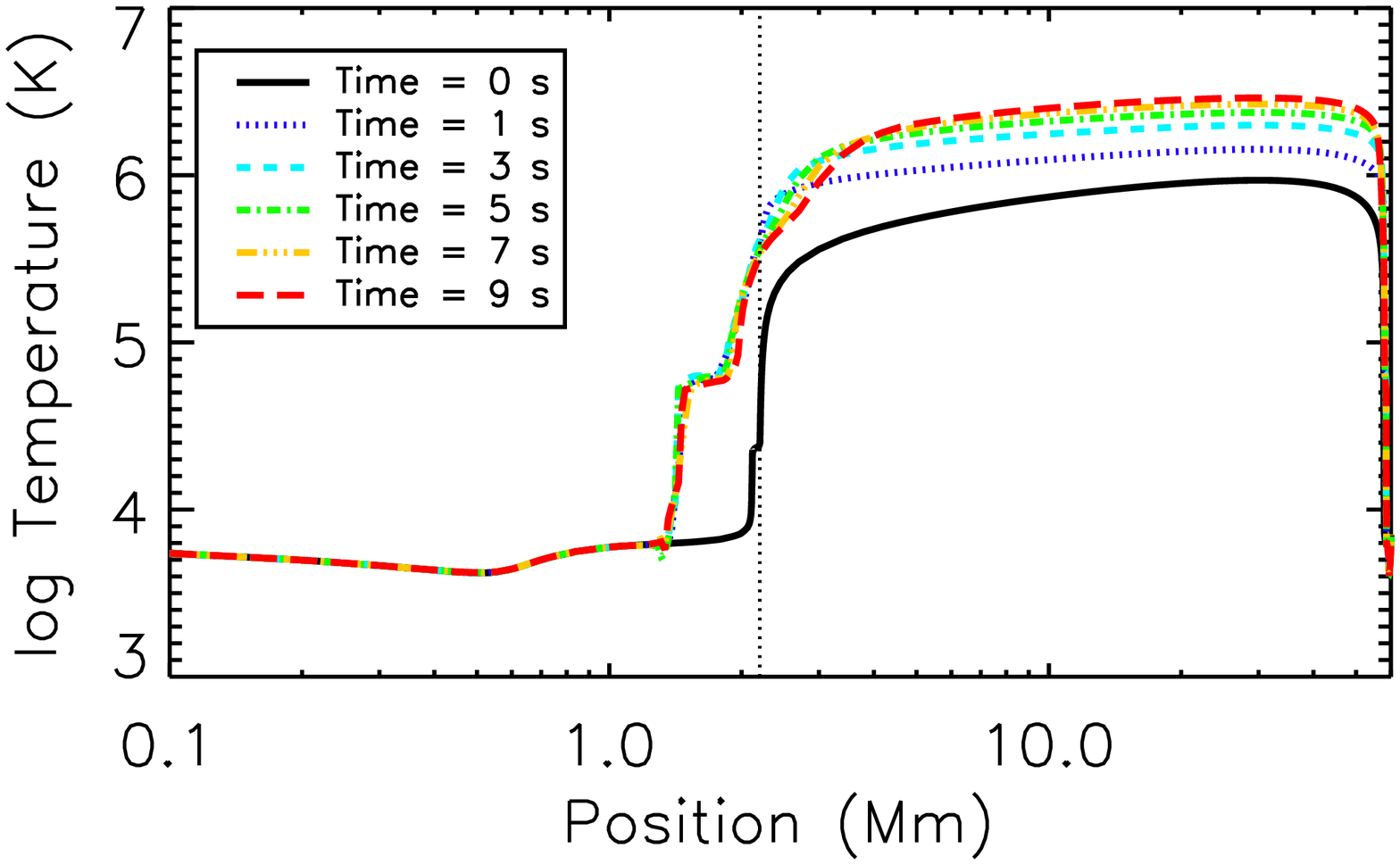}
\end{minipage}
\begin{minipage}[b]{0.33\linewidth}
\centering
\includegraphics[width=\textwidth]{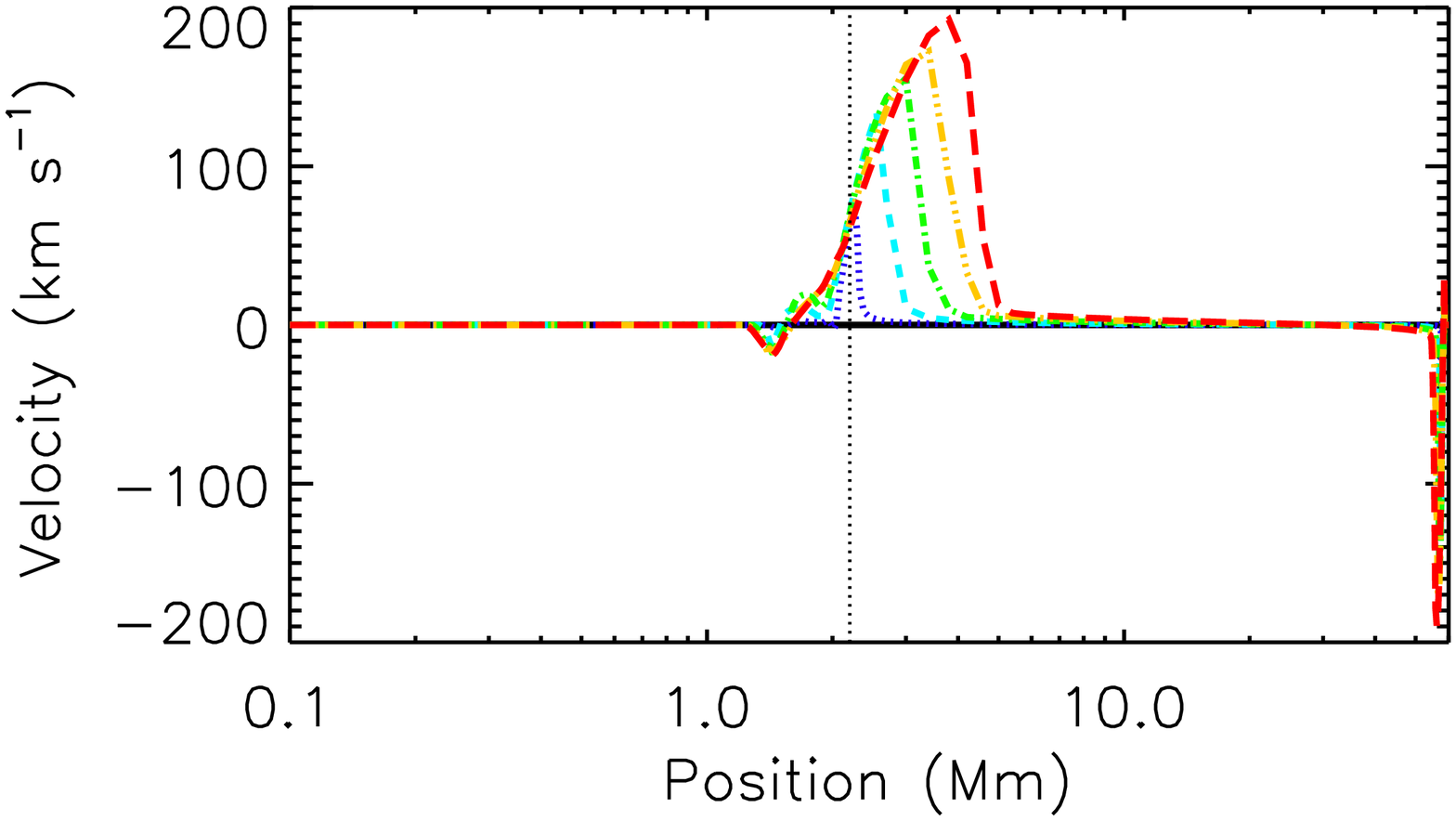}
\end{minipage}
\begin{minipage}[b]{0.33\linewidth}
\centering
\includegraphics[width=\textwidth]{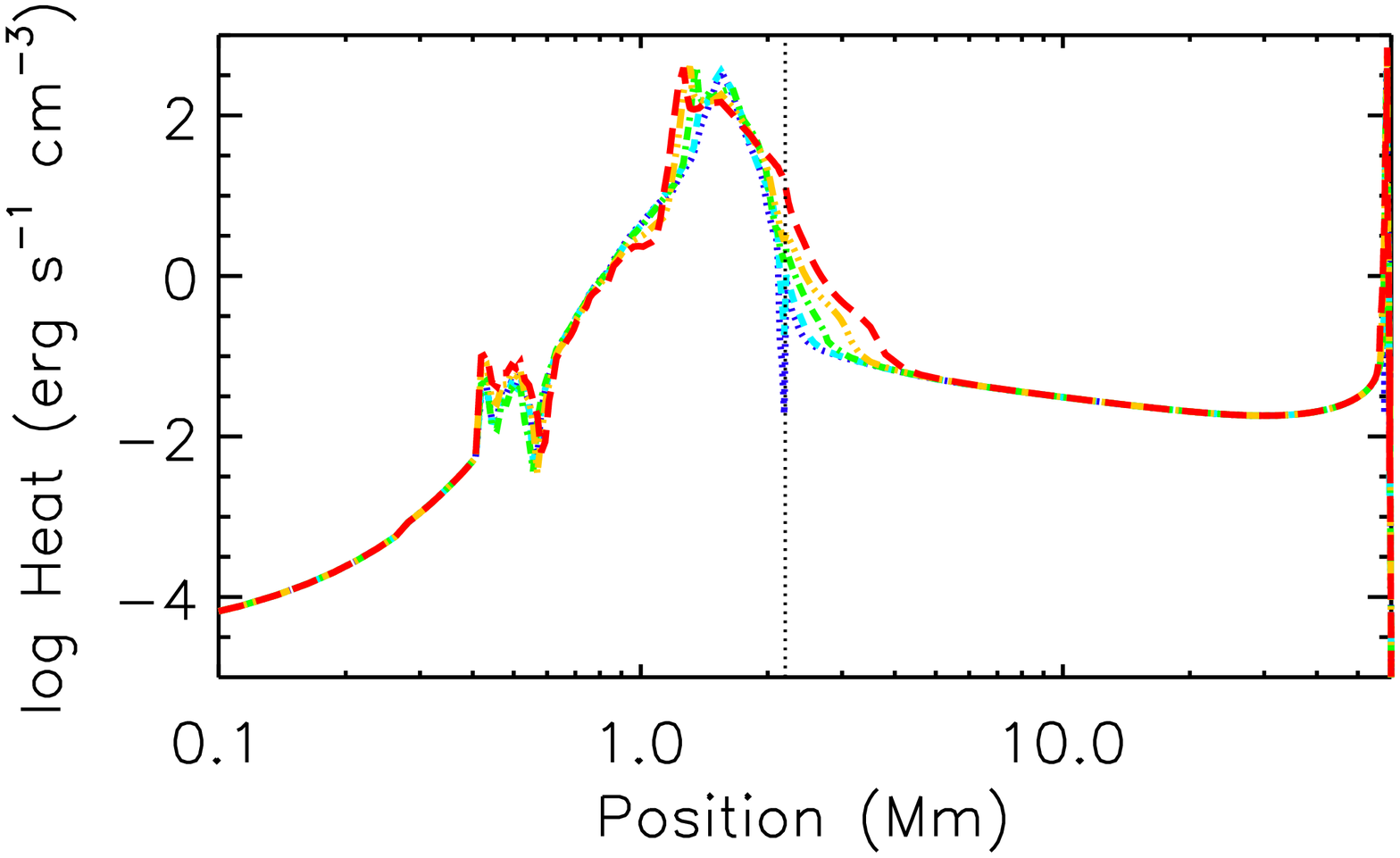}
\end{minipage}
\begin{minipage}[b]{0.33\linewidth}
\centering
\includegraphics[width=\textwidth]{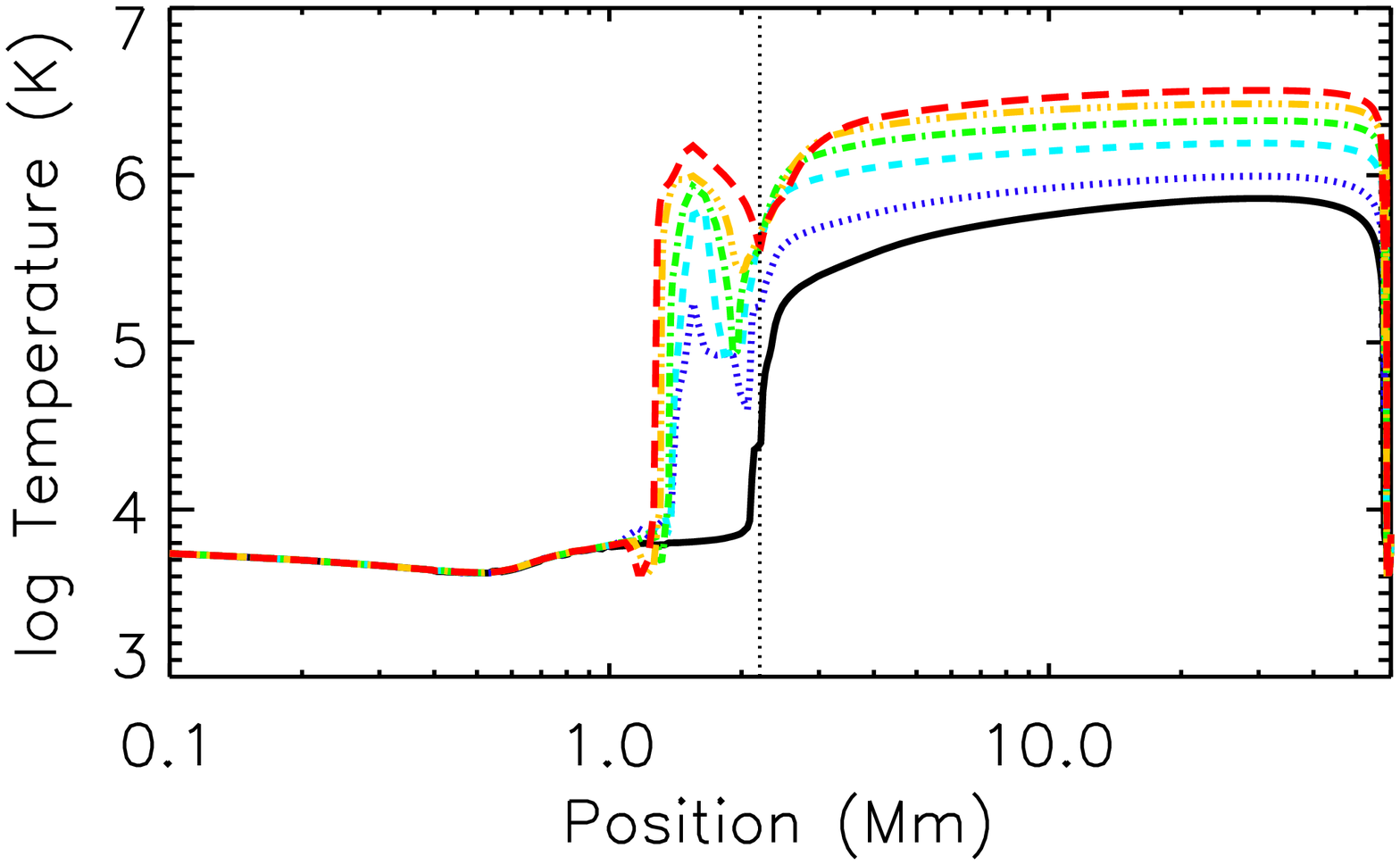}
\end{minipage}
\begin{minipage}[b]{0.33\linewidth}
\centering
\includegraphics[width=\textwidth]{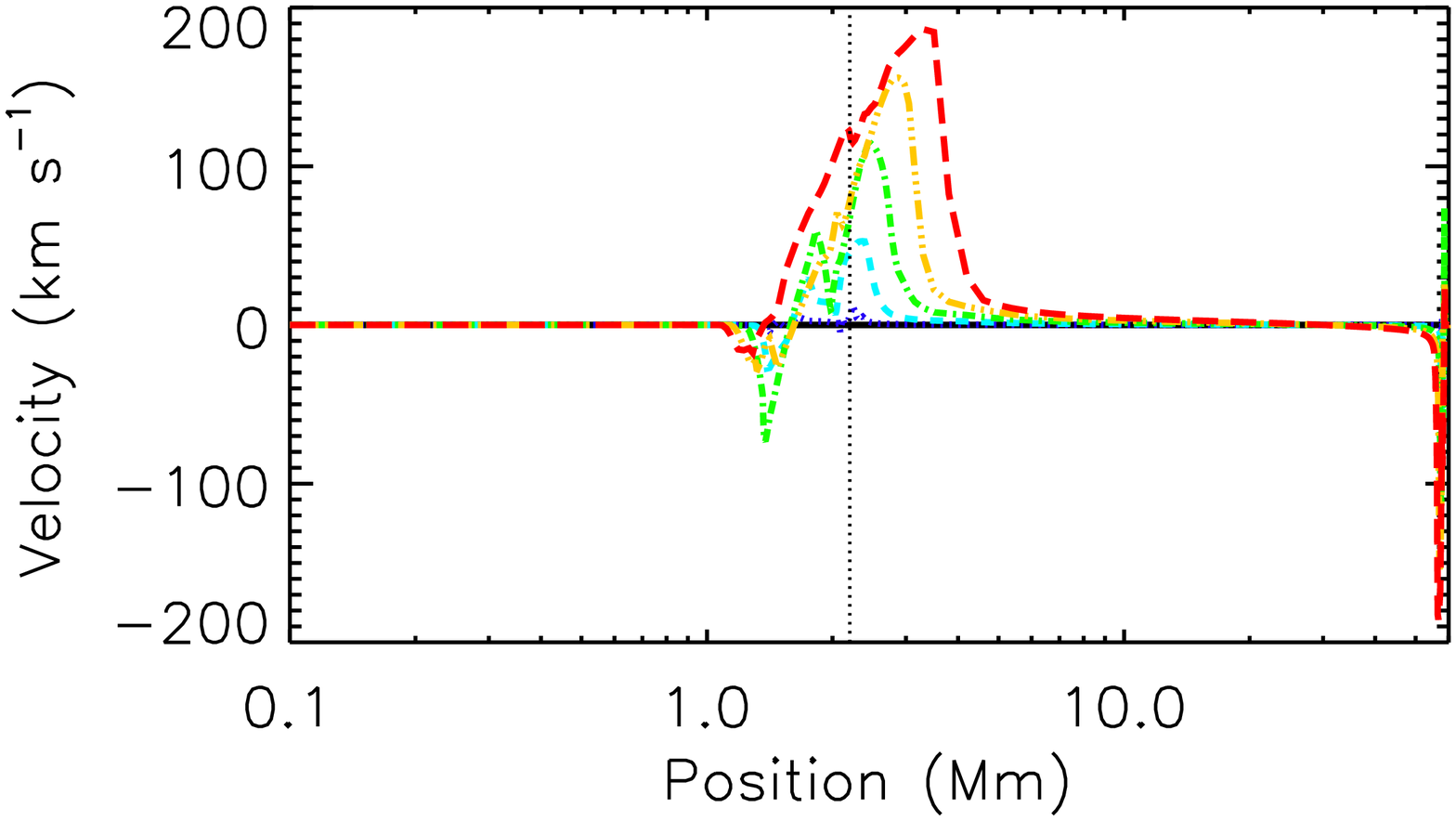}
\end{minipage}
\caption{A comparison between three scenarios: heating by Alfv\'enic waves with $f = 10$\,Hz, $k_{x,a} = 10^{-5}$\,cm$^{-1}$ (top row), $k_{x,a} = 4 \times 10^{-4}$\,cm$^{-1}$ (middle row), and an electron beam with $E_{c} = 20$\,keV (bottom row).  At a few selected time periods, the first column shows the total energy deposited in the loop, the second column the electron temperature, and the third column the bulk flow velocity.  Note that the x-axis is logarithmic in these plots, with the loop apex at 30 Mm.  The velocity plots define right-moving flows as positive, left-moving negative.  The dashed vertical lines mark the initial transition region boundary.} 
\label{heatingscenarios}
\end{figure*}

We consider first a simulation with wave heating for $k_{x,a} = 10^{-5}$\,cm$^{-1}$, $f = 10$\,Hz, and $k_{x}$ scaling linearly with the magnetic field.  The top row of Figure \ref{heatingscenarios} shows the atmospheric response to 10 seconds of heating, with an initial Poynting flux of $10^{10}$\,erg s$^{-1}$ cm$^{-2}$ (note the x-axis is logarithmic and extends from foot-point to foot-point).  With increasing depth into the chromosphere, the density rises and the ionization fraction falls, increasing the effectiveness of ion-neutral friction.  The temperature minimum region is strongly heated, with a non-negligible amount of heating in the upper chromosphere.  As the temperature rises slowly, so does the pressure, causing a gentle evaporation to form in the transition region, reaching about 50\,km s$^{-1}$ in 10 seconds (the plot defines right-moving flows as positive, left-moving as negative).  The corona is essentially unaffected by these waves, as they propagate through with only minimal damping.  

Contrast this now with a simulation that has a much higher perpendicular wave number, $k_{x,a} = 4 \times 10^{-4}$\,cm$^{-1}$, but otherwise equal properties, shown in the middle row of Figure \ref{heatingscenarios}.  Due to the increase in the wave number, the waves are strongly damped in the upper atmosphere by collisions between ions and electrons.  The upper chromosphere is strongly heated, as the Poynting flux sharply decreases across this layer so that only minuscule amounts of energy are carried to the temperature minimum region.  Deeper in the chromosphere, the heating falls off as the Poynting flux dissipates, and is not a smooth function primarily due to changes in the ionization state of the plasma.  The pressure increase in the upper chromosphere is sharp enough that material explosively evaporates, reaching over 200\,km s$^{-1}$ in 10 seconds of heating.  If the heating were sustained, the density in the corona would increase significantly, causing brightening across the extreme ultraviolet and soft X-rays characteristic of flares.  

For comparison, consider heating by an electron beam in the CTTM (using the model of \citealt{emslie1978} with a sharp cut-off).  We adopt a low-energy cut-off $E_{c} = 20$\,keV, spectral index $\delta = 5$, and energy flux $F_{0} = 10^{10}$\,erg s$^{-1}$ cm$^{-2}$ (equal to the Poynting flux considered), shown in the bottom row of Figure \ref{heatingscenarios}.  Compared to the previous simulation, slightly more energy is deposited in the corona as the electrons collide with ambient particles there, and a comparable amount of energy in the chromosphere.  The temperature in the upper chromosphere and corona rises slightly higher than in the previous simulation, while the evaporation again reaches about $200$\,km s$^{-1}$.  The atmospheric response is nearly identical, and without a direct measure of the energy input, would be difficult to distinguish observationally.  

The explanation for the different wave-driven behaviors -- namely gentle versus explosive evaporation -- is straight-forward.  As seen in Equation \ref{dampinglength}, the damping length is shorter for higher frequencies (which increase the damping by perpendicular resistivity) or higher perpendicular wave numbers (which increase the damping by parallel resistivity).  In this regard, our simulation results are consistent with the findings of \citet{emslie1982}, who showed that waves with higher frequency or perpendicular wave number do not penetrate the deep chromosphere because they dissipate higher in the atmosphere.  

\begin{figure*}
\begin{minipage}[b]{0.33\linewidth}
\centering
\includegraphics[width=\textwidth]{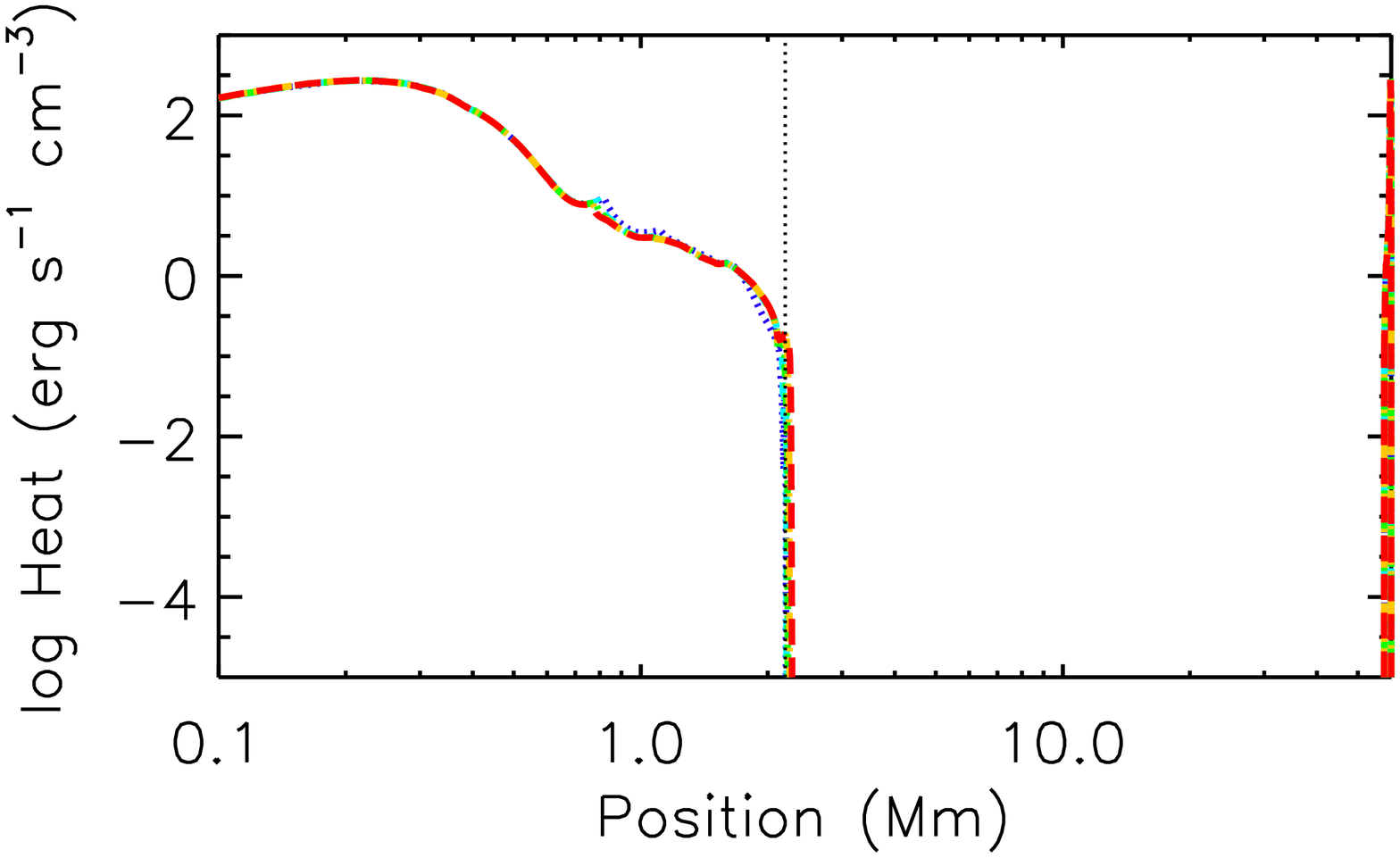}
\end{minipage}
\begin{minipage}[b]{0.33\linewidth}
\centering
\includegraphics[width=\textwidth]{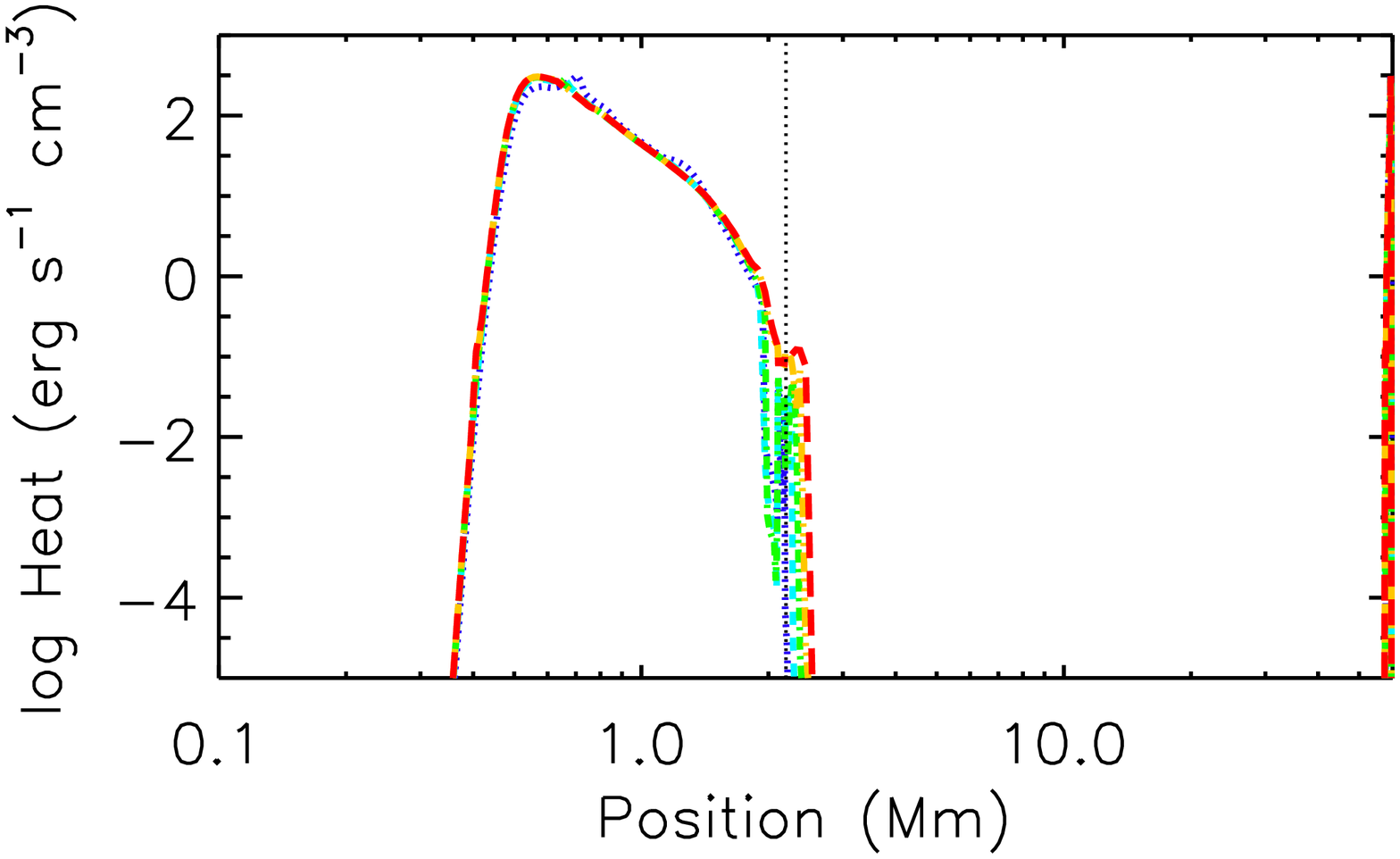}
\end{minipage}
\begin{minipage}[b]{0.33\linewidth}
\centering
\includegraphics[width=\textwidth]{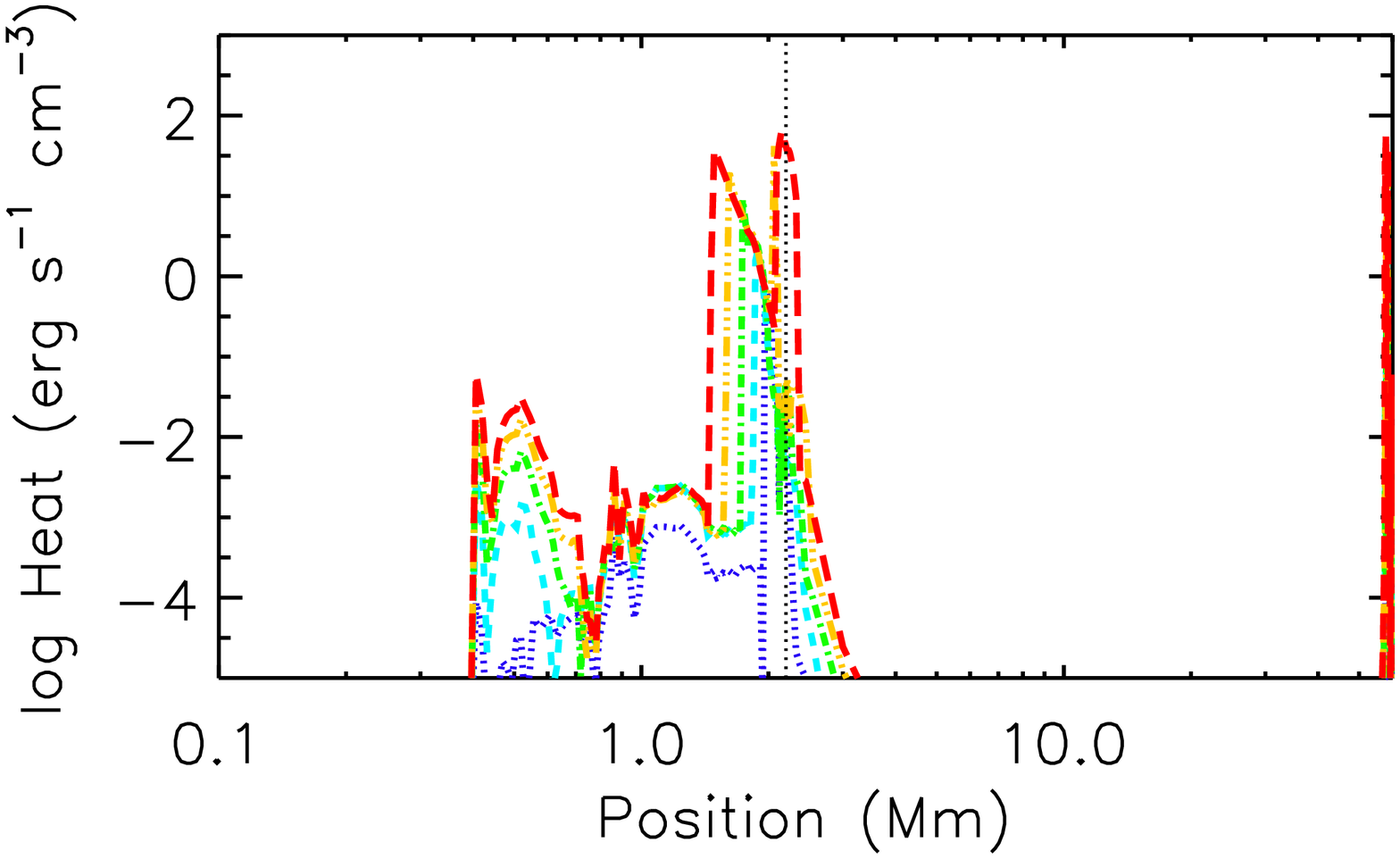}
\end{minipage}
\begin{minipage}[b]{0.33\linewidth}
\centering
\includegraphics[width=\textwidth]{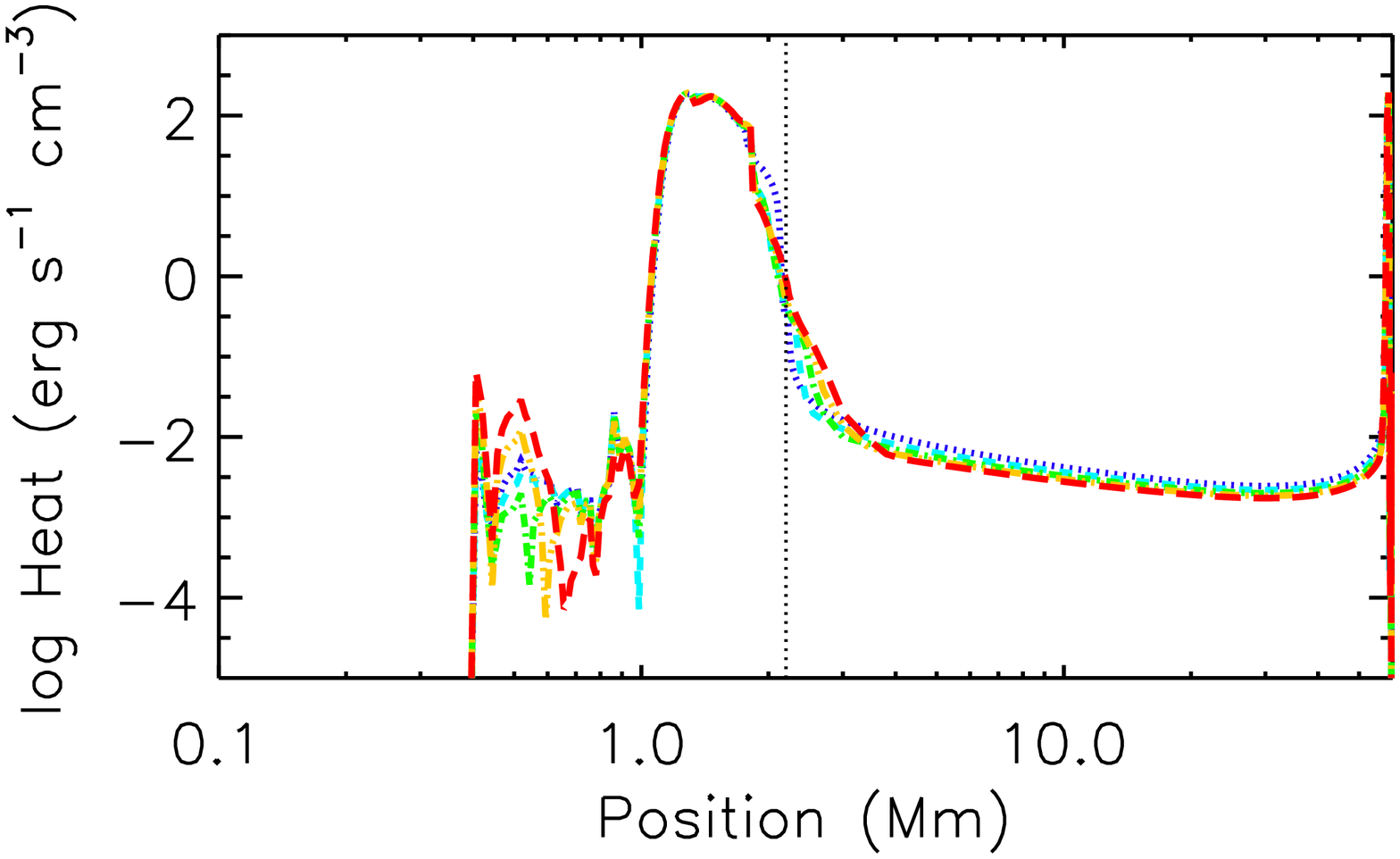}
\end{minipage}
\begin{minipage}[b]{0.33\linewidth}
\centering
\includegraphics[width=\textwidth]{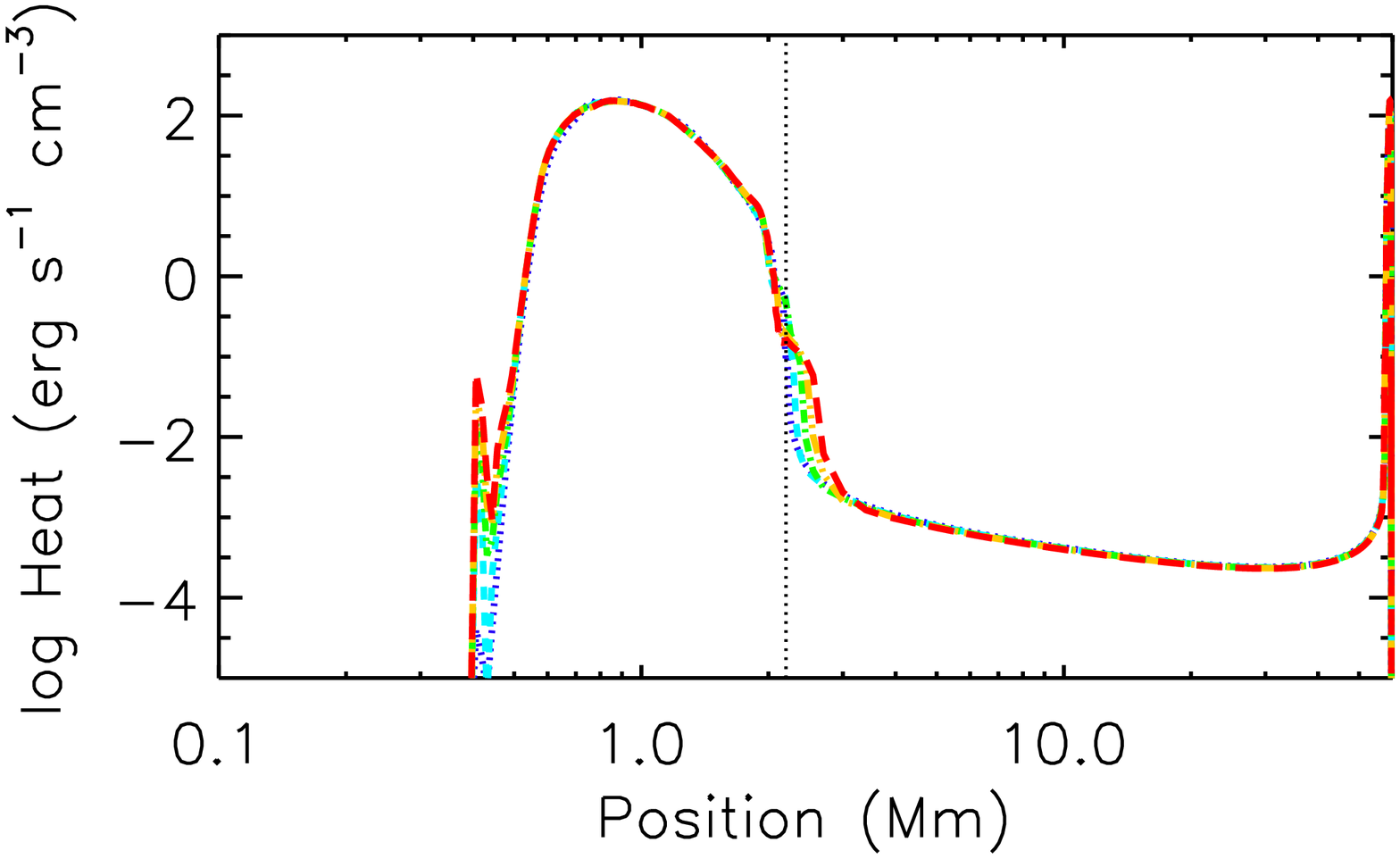}
\end{minipage}
\begin{minipage}[b]{0.33\linewidth}
\centering
\includegraphics[width=\textwidth]{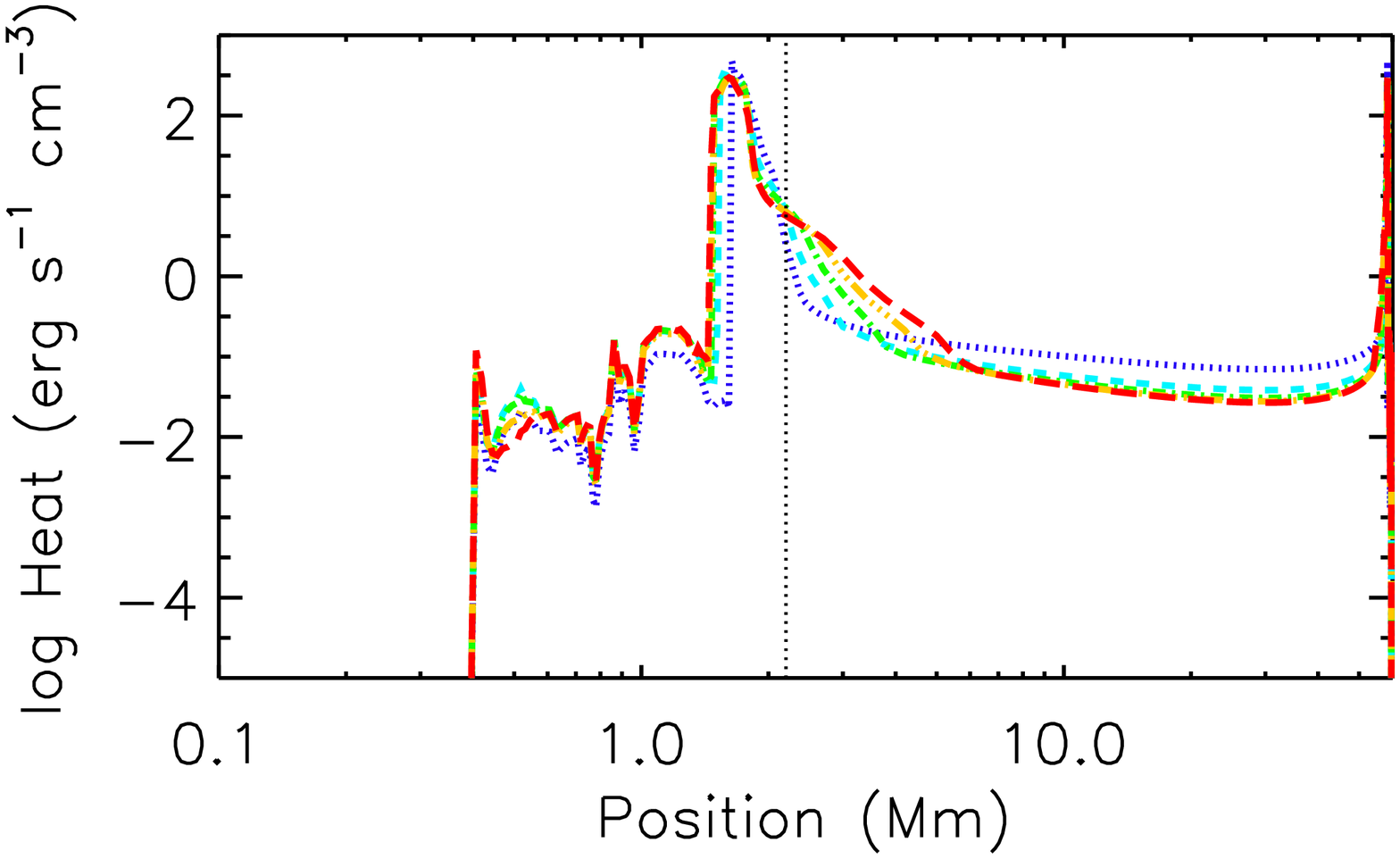}
\end{minipage}
\caption{Energy deposition plots for wave heating with various parameters.  The top row shows three simulations with $k_{x} = 0$, {\it i.e.} without perpendicular damping, and $f = 1, 10, 100$\,Hz, respectively.  The bottom left plot shows $k_{x,a} = 10^{-4}$\,cm$^{-1}$ and $f = 10$\,Hz.  The last two plots are the same as the wave heating simulations in Figure \ref{heatingscenarios}, except that their wave number scales as $k_{x}(z) = k_{x,a} \sqrt{B(z)/B_{a}}$.  Note that the x-axis is logarithmic, as before.}
\label{parameters}
\end{figure*}

The top row of Figure \ref{parameters} explicitly shows the change with wave frequency, with plots of heating from three simulations with frequencies of 1, 10, 100\,Hz, all for the same wave number, $k_{x} = 0$.  Note that without damping from $k_{x}$, waves barely heat the corona.  Similarly, the effect of increasing $k_{x}$ for a fixed frequency is seen by comparing the top center plot in Figure \ref{parameters} ($k_{x} = 0$), the top row of Figure \ref{heatingscenarios} ($k_{x,a} = 10^{-5}$ cm$^{-1}$), the bottom left panel of Figure \ref{parameters} ($k_{x,a} = 10^{-4}$ cm$^{-1}$), and the middle row of Figure \ref{heatingscenarios} ($k_{x,a} = 4 \times 10^{-4}$ cm$^{-1}$).

Finally, supposing that $k_{x}(z)$ does not vary linearly, but as the square root of the magnetic field ($k_{x}(z) = k_{x,a} \sqrt{B(z)/B_{a}}$), the last two plots of Figure \ref{parameters} repeat the simulations in Figure \ref{heatingscenarios}, with otherwise identical parameters.  The main difference is that because the dependence on the magnetic field is reduced, the wave number is higher at larger heights, so that the waves dissipate more of their energy there.  

\section{Conclusions}
\label{sec:conclusions}

The main conclusions are as follows:

\begin{enumerate}
\item[(1)] Flare-generated Alfv\'enic waves can heat not only the temperature minimum region, but also the upper chromosphere and corona.  
\item[(2)] Heating by dissipation of Alfv\'enic waves can be very similar to heating by electron beams ({\it e.g.} \citealt{brown1971, emslie1978}).  
\item[(3)] Since the upper chromosphere can be strongly heated, Alfv\'enic waves can cause explosive evaporation.  As with beam heating models, the pressure in the chromosphere rises sharply, causing a rapid expansion of material.  
\end{enumerate}

Ion-neutral friction damps down-going waves at the temperature minimum region (as found by \citealt{emslie1982} and \citealt{russell2013}) and it becomes important in the upper atmosphere for high frequencies.  The frequencies required to heat the upper chromosphere this way are sensitive to the field strength, but in our simulations 10\,Hz waves produced a pronounced heating and 100\,Hz waves were almost entirely absorbed there.  Millisecond spikes in radio and HXRs \citep{kiplinger1983,benz1986} show that flares produce such frequencies, and {\it in situ} observations of magnetospheric reconnection show generation of Alfv\'enic waves with high frequencies \citep{keiling2009}.  It is therefore credible that Alfv\'enic waves excited during flares would also include a component that heats the upper chromosphere by ion-neutral friction.  

Parallel resistivity can also have a significant role if the incident waves have fine structure perpendicular to the magnetic field (not considered by \citealt{russell2013}).  In our simulations, incident waves with $k_{x,a} \ge 10^{-4}$\,cm$^{-1}$ produced considerable heating in the upper chromosphere.  This wave number corresponds to a scale of 600\,m, which is two orders of magnitude larger than the coronal proton inertial length (assuming $n \approx 10^{9}$ cm$^{-3}$) and at least two orders of magnitude larger than the proton Larmor radius (assuming $T \lesssim 4 \times 10^{7}$\,K and $B \gtrsim 10$\,G).  If Alfv\'enic waves are produced in the corona by 3D collisionless reconnection, then it is reasonable to expect they will inherit their scales from the reconnection dynamics, which produce flux ropes with scales of tens of ion inertial lengths \citep{daughton2011}.  On the other hand, if waves are produced on larger scales, various coronal processes act to reduce perpendicular scales, for example: magnetic convergence and phase mixing \citep{russell2013b}; Kelvin-Helmholtz and tearing instabilities \citep{chaston2010}; Alfv\'enic cascades \citep{goldreich1995}; and mapping along braided magnetic fields \citep{pontin2015}.  Thus, there are grounds to expect that part of the wave power produced by flares would arrive at the chromosphere with scales that lead to resistive damping by electron collisions.  

Since the heating and evaporation is similar to electron beams, can they be distinguished observationally?  EUV and SXR emissions, primarily dependent on density and temperature changes, must also be similar.  A non-thermal HXR burst indicates the presence of accelerated electrons, but if waves can accelerate electrons in the chromosphere or low corona \citep{fletcher2008,melrose2014}, or if waves travel along the same flux tubes as electrons, the presence of an HXR burst alone does not rule out Alfv\'enic wave heating.  The similarity of the heating signatures is particularly problematic for studies of nanoflares, where HXR emission, if present, is too faint to be detected. For example, \citet{testa2014} recently investigated chromospheric heating during nanoflares and found that IRIS observations are consistent with heat input by nonthermal particles; our results suggest that similar signatures could also be produced by wave heating.  New HXR instruments such as FOXSI \citep{krucker2014} and NuStar \citep{harrison2013}, with improved sensitivity and spatial resolution, may help resolve this.     

It seems possible that Alfv\'enic waves can play an important role in flare heating.  It is undeniable, however, that there are many electrons being accelerated in flares, producing strong HXR bursts, which are well correlated with the rise in SXR emission \citep{dennis1993}.  Therefore, future work needs to further establish the viability of this heating mechanism, but also to what extent it operates {\it in tandem} with electron beams, and how energy might be partitioned between them.

\vspace{5mm}
\noindent This research was performed while JWR held an NRC Research Associateship award at the US Naval Research Laboratory with the support of NASA.  AJBR was supported by the U.K. Science and Technology Facilities Council under grant ST/K000993/1 to the University of Dundee.  The authors benefited from participation in the International Space Science Institute team on ``Magnetic Waves in Solar Flares: Beyond the `Standard' Flare Model," led by Alex Russell and Lyndsay Fletcher.  We acknowledge particularly helpful comments about resistive damping of waves with large $k_{\perp}$ from James Leake and Peter Damiano, which influenced this work.  We thank Martin Laming and Graham Kerr for comments that helped to improve this paper.


\begin{thebibliography}{}
\bibitem[Alfv\'en(1942)]{alfven1942} Alfv\'en, H.\ 1942, \nat, 150, 405
\bibitem[Benz(1986)]{benz1986} Benz, A.O.\ 1986, \solphys, 104, 99
\bibitem[Birn et al.(2009)]{birn2009} Birn, S.J., Fletcher, L., Hesse, M., \& Neukirch, T.\ 2009, \apj, 695, 1151
\bibitem[Bradshaw \& Cargill(2013)]{bradshaw2013} Bradshaw, S.J., \& Cargill, P.J.\ 2013, \apj, 770, 12
\bibitem[Bradshaw \& Mason(2003)]{bradshaw2003} Bradshaw, S.J., \& Mason, H.E.\ 2003, \aap, 401, 699
\bibitem[Brown(1971)]{brown1971} Brown, J.C.\ 1971, \solphys, 18, 489
\bibitem[Brown et al.(2009)]{brown2009} Brown, J.C., Turkmani, R., Kontar, E.P., et al.\ 2009, \aap, 508, 993
\bibitem[Carlsson \& Leenaarts(2012)]{carlsson2012} Carlsson, M., \& Leenaarts, J.\ 2012, \aap, 539, 39
\bibitem[Carmichael(1964)]{carmichael1964} Carmichael, H.\ 1964, NASSP, 50, 451
\bibitem[Chaston \& Seki(2010)]{chaston2010} Chaston, C.C., \& Seki, K.\ 2010, \jgr, 115, A11221
\bibitem[Daughton et al.(2011)]{daughton2011} Daughton, W., Roytershteyn, V., Karimabadi, H., et al.\ 2011, Nature Phys., 7, 539
\bibitem[De Pontieu et al.(2001)]{depontieu2001} De Pontieu, B., Martens, P.C.H., \& Hudson, H.S.\ 2001, \apj, 558, 859
\bibitem[Del Zanna et al.(2015)]{delzanna2015} Del Zanna, G., Dere, K.P., Young, P.R., et al.\ 2015, \aap, 582, 56
\bibitem[Dennis \& Zarro(1993)]{dennis1993} Dennis, B.R., \& Zarro, D.M.\ 1993, \solphys, 146, 177
\bibitem[Dere et al.(1997)]{dere1997} Dere, K.P., Landi, E., Mason, H.E., et al.\ 1997, \aaps, 125, 149
\bibitem[Emslie(1978)]{emslie1978} Emslie, A.G.\ 1978, \apj, 224, 241
\bibitem[Emslie \& Machado(1979)]{emslie1979} Emslie, A.G., \& Machado, M.E.\ 1979, \solphys, 64, 129
\bibitem[Emslie \& Sturrock(1982)]{emslie1982} Emslie, A.G., \& Sturrock, P.A.\ 1982, \solphys, 80, 99
\bibitem[Fletcher et al.(2011)]{fletcher2011} Fletcher, L., Dennis, B.R., Hudson, H.S., et al.\ 2011, \ssr, 159, 19
\bibitem[Fletcher \& Hudson(2008)]{fletcher2008} Fletcher, L., \& Hudson, H.S.\ 2008, \apj, 675, 1645
\bibitem[Goldreich \& Sridhar(1995)]{goldreich1995} Goldreich, P., \& Sridhar, S.\ 1995, \apj, 438, 763
\bibitem[Harrison et al.(2013)]{harrison2013} Harrison, F.A., Craig, W.W., Christensen, F.E. et al.\ 2013, \apj, 770, 103
\bibitem[Hirayama(1974)]{hirayama1974} Hirayama, T.\ 1974, \solphys, 34, 323
\bibitem[Holman et al.(2003)]{holman2003} Holman, G.D., Sui, L., Schwartz, R.A., \& Emslie, A.G.\ 2003, \apjl, 595, L97
\bibitem[Keiling(2009)]{keiling2009} Keiling, A.\ 2009, \ssr, 142, 73
\bibitem[Khodachenko et al.(2004)]{khodachenko2004} Khodachenko, M.L., Arber, T.D., Rucker, H.O., \& Hanslmeier, A.\ 2004, \aap, 422, 1073
\bibitem[Kigure et al.(2010)]{kigure2010} Kigure, H., Takahashi, K., Shibata, K., et al.\ 2010, \pasj, 62, 993
\bibitem[Kiplinger et al.(1983)]{kiplinger1983} Kiplinger, A.L., Dennis, B.R., Frost, K.J., et al.\ 1983, \apjl, 265, L99
\bibitem[Klimchuk(2006)]{klimchuk2006} Klimchuk, J.A.\ 2006, \solphys, 234, 41
\bibitem[Kopp \& Pneuman(1976)]{kopp1976} Kopp, R.A., \& Pneuman, G.W.\ 1976, \solphys, 50, 85
\bibitem[Krucker et al.(2014)]{krucker2014} Krucker, S., Christe, S., Glesener, L., et al.\ 2014, \apjl, 793, L32
\bibitem[Krucker et al.(2011)]{krucker2011} Krucker, S., Hudson, H.S., Jeffrey, N.L.S., et al.\ 2011, \apjl, 739, L96
\bibitem[Laming(2004)]{laming2004} Laming, J.M.\ 2004, \apj, 614, 1063
\bibitem[Laming(2015)]{laming2015} Laming, J.M.\ 2015, LRSP, 12, 2
\bibitem[Leake et al.(2014)]{leake2014} Leake, J.E., DeVore, C.R., \& Thayer, J.P.\ 2014, \ssr, 184, 107
\bibitem[Machado et al.(1978)]{machado1978} Machado, M.E., Emslie, A.G., \& Brown, J.C.\ 1978, \solphys, 58, 363
\bibitem[Melrose \& Wheatland(2014)]{melrose2014} Melrose, D.B., \& Wheatland, M.S.\ 2014, \solphys, 289, 881
\bibitem[McIntosh et al.(2011)]{mcintosh2011} McIntosh, S.W., de Pontieu, B., Carlsson, M., et al.\ 2011, \nat, 475, 477
\bibitem[Piddington(1956)]{piddington1956} Piddington, J.H.\ 1956, \mnras, 116, 314
\bibitem[Pontin \& Hornig(2015)]{pontin2015} Pontin, D.I., \& Hornig, G.\ 2015, \apj, 805, 47
\bibitem[Reep et al.(2013)]{reep2013} Reep, J.W., Bradshaw, S.J., \& McAteer, R.T.J.\ 2013, \apj, 778, 76
\bibitem[Russell \& Fletcher(2013)]{russell2013} Russell, A.J.B., \& Fletcher, L.\ 2013, \apj, 765, 81
\bibitem[Russell \& Stackhouse(2013)]{russell2013b} Russell, A.J.B., \& Stackhouse, D.J.\ 2013, \aap, 558, 76
\bibitem[Soler et al.(2013)]{soler2013} Soler, R., Carbonell, M., Ballester, J.L., \& Terradas, J.\ 2013, \apj, 767, 171
\bibitem[Soler et al.(2015)]{soler2015} Soler, R., Ballester, J.L., \& Zaqarashvili, T.V.\ 2015, \aap, 573, 79
\bibitem[Sturrock(1966)]{sturrock1966} Sturrock, P.A.\ 1966, \nat, 211, 697
\bibitem[Testa et al.(2014)]{testa2014} Testa, P., De Pontieu, B., Allred, J., et al.\ 2014, Science, 346, 1255724
\bibitem[Tomczyk et al.(2007)]{tomczyk2007} Tomczyk, S., McIntosh, S.W., Keil, S.L., et al.\ 2007, Science, 317, 1192
\bibitem[Varady et al.(2014)]{varady2014} Varady, M., Karlick\'{y}, M., Moravec, Z., \& Ka\v{s}parov\'{a}, J.\ 2014, \aap, 563, 51
\bibitem[Vernazza et al.(1981)]{vernazza1981} Vernazza, J.E., Avrett, E.H., \& Loeser, R.\ 1981, \apjs, 45, 635

\end{thebibliography}
\end{document}